\documentclass[aps,prl,twocolumn,showpacs,superscriptaddress]{revtex4-2}
\usepackage{graphicx}
\usepackage{dcolumn}
\usepackage{bm}
\usepackage{hyperref}
\usepackage[T1]{fontenc}
\usepackage[utf8]{inputenc}
\usepackage[english]{babel}
\usepackage{units}
\usepackage{mathtools}
\hypersetup{colorlinks=true,citecolor={blue},linkcolor={blue},urlcolor={blue}}
\usepackage{amsmath}
\usepackage{empheq}
\usepackage{mathrsfs}
\usepackage{amssymb}
\usepackage{soul}
\usepackage{upgreek}
\usepackage{xcolor}

\begin{document}


\title{Vortex clusters in a stirred polariton condensate}

\author{I. Gnusov}
\email[I. Gnusov]{Ivan.Gnusov@skoltech.ru}
\address{Skolkovo Institute of Science and Technology, Moscow, Territory of innovation center “Skolkovo”,
Bolshoy Boulevard 30, bld. 1, 121205, Russia.}

\author{S. Harrison}
\address{School of Physics and Astronomy, University of Southampton,  Southampton, SO17 1BJ, UK.}

\author{S. Alyatkin}
\address{Skolkovo Institute of Science and Technology, Moscow, Territory of innovation center “Skolkovo”,
Bolshoy Boulevard 30, bld. 1, 121205, Russia.}

\author{K. Sitnik}
\address{Skolkovo Institute of Science and Technology, Moscow, Territory of innovation center “Skolkovo”,
Bolshoy Boulevard 30, bld. 1, 121205, Russia.}

\author{H. Sigur{\dh}sson}
\affiliation{Institute of Experimental Physics, Faculty of Physics, University of Warsaw, ul.~Pasteura 5, PL-02-093 Warsaw, Poland}
\affiliation{Science Institute, University of Iceland, Dunhagi 3, IS-107 Reykjavik, Iceland} 


\author{P. G. Lagoudakis}
\address{Skolkovo Institute of Science and Technology, Moscow, Territory of innovation center “Skolkovo”,
Bolshoy Boulevard 30, bld. 1, 121205, Russia.}
\address{School of Physics and Astronomy, University of Southampton,  Southampton, SO17 1BJ, UK.}

\begin{abstract}

The response of superfluids to the external rotation, evidenced by emergence of quantised vortices, distinguishes them from conventional fluids. In this work, we demonstrate that the number of vortices in a stirred polariton condensate depends on the characteristic size of the employed rotating potential induced by the nonresonant laser excitation. For smaller sizes, a single vortex with a topological charge of $\pm1$ corresponding to the stirring direction is formed. However, for larger optical traps, clusters of two or three co-rotating vortices appear in the narrow range of GHz stirring speed.


\end{abstract}

\maketitle 

The emergence of quantised vorticity in rotating quantum fluids is a well-known hallmark of superfluidity. The classical experiments with stirred superfluid Helium~\cite{hevortex, HePackard_PhysRevA.6.799} and Bose-Einstein condensates in ultracold atomic gases~\cite{bec_laser_prl1999,bec2component,bec_review} showcased that vortices appear above some critical rotation speed (frequency) and their number grows with increased stirring. In such systems the vortices self-arrange to form clusters with the energy-favourable triangular geometry~\cite{hevortex,bec2component,Butts1999}. Only quite recently was quantised vorticity reported in a completely different platform, a nonresonantly optically stirred condensatee of exciton-polaritons~\cite{rotatingbucket,fraser2023}, opening a new pathway to explore optical vorticity in dynamically driven condensed matter~\cite{Quinteiro_RMP2022, Yulin_Arxiv2023}. 

Exciton-polaritons (from here on \textit{polaritons}) are bosonic quasiparticles that arise due to the strong coupling of the excitons and photons in semiconductor microcavities~\cite{kavokin_microcavities_2007}. Optical malleability and characterisation through the leaky cavity photon mode make polaritons a perfect playground to study many-body physics and macroscopic coherent phenomena driven far from equilibrium. A polariton condensate ~\cite{ huideng_review, kasprzak_bose-einstein_2006} is a coherent state described by a macroscopic polariton wavefunction which, depending on the excitation conditions, can display superfluid behaviour~\cite{Amo2009superfluid,Lerario2017}. From the first experimental observation of quantised polariton vortices in 2008~\cite{Lagoudakis2008firtsv}, their study has been thriving to date with reports on half-quantised vortices~\cite{Lagoudakis2009_halfquantised, Dominici_SciAdv2015}, optically trapped vortices~\cite{chiral_lens_prl, Ma2020_vortex_swotching, Berger_PRB2020}, vortex-antivortex pairs~\cite{Roumpos2010_vortex-antivortex}, oscillating vortex clusters~\cite{sitnik_2022}, high charge vortices~\cite{Sedov_PRR2021, Cookson_NatComm2021}, chains~\cite{Boulier_SciRep2015, Gao_vortex_chain} and lattices~\cite{resonantvortexlattice,nonresonant_lattice_spots,sergey_vortex_lattice, Wang_NaScRev2022}, and turbulence~\cite{Nardin2011_flowdeff, Caputo_NatPhot2019, Panico2023_alotvortices}. Except for the fundamental interests, these studies are fueled by intriguing proposals on the usage of circulating polariton currents for both simulating spin systems~\cite{sergey_vortex_lattice, Harrison_OptMatExpr2023} and unconventional computing purposes~\cite{Sigurdsson_PRB2014, Kavokin2022}. 

Even though the first observation of polariton superfluidity was reported almost 15 years ago~\cite{Amo2009superfluid}, it was not until recently when the signature experiments with stirred polariton fluids using incoherent laser light were implemented~\cite{rotatingbucket, fraser2023}. The reason for this delay lies in the extremely high and, at first glance, experimentally elusive GHz rotation speed~\cite{rotatingbucket} required to stir the polariton condensate. Interestingly, the driven-dissipative nature of the polariton fluid leads to significant differences in its rotation dynamics compared to the conventional superfluids~\cite{hevortex,bec_review}. The most intriguing and counter-intuitive one is that to date, only a single vortex has been observed in the stirred polariton condensate~\cite{rotatingbucket, fraser2023}, even though the numerical simulations predicted the formation of a cluster of co-rotating vortices~\cite{fraser2023}. 

Here, we demonstrate the formation of clusters of up to three co-rotating vortices in a nonresonantly optically stirred polariton condensate. The vortex clusters appear only in a narrow range of GHz stirring frequencies, with the number of vortices determined by the dimensions of the time-periodic confining potential. This is corroborated through measurements on the statistical occurrence of the vortices as a function of rotation frequency. Furthermore, at a fixed stirring frequency, we reveal a transition from two co-rotating vortices to a single vortex with increasing excitation laser power. We stress that all vortices in externally driven polariton fluids have the same topological charge dictated only by the stirring direction. Our experimental observations are supported by numerical simulations. 

\begin{figure}[hbt]
    \centering
    \includegraphics[width=0.95\columnwidth]{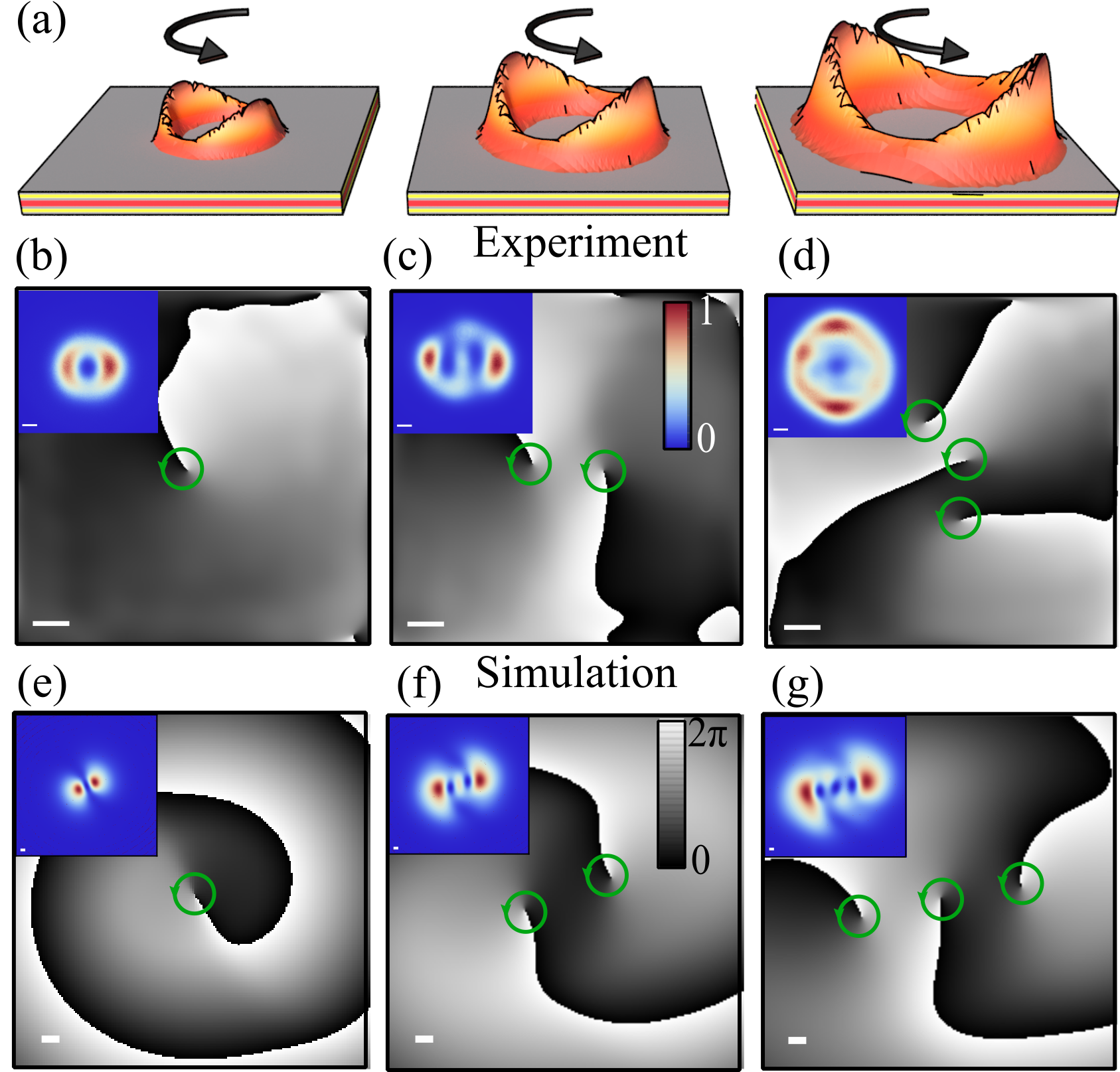}
    \caption{(a) The schematic demonstrating three rotating optical traps (yellow to orange colour scale) of different sizes. Experimentally measured real-space phase distribution of the condensate in the rotating trap with diameter of (b) 14 $\mu$m, (c) 15.4 $\mu$m, (d) 17.5 $\mu$m. The insets in (b-d) show the time-integrated condensate intensity distribution. (e-g) Corresponding simulated instantaneous real-space phases with insets in panels, depicting the snapshots of the condensate intensity distribution. The scale bars in (b-g) equals 2 $\mu$m. }
    \label{fig1}
\end{figure}

The experiments are carried out using an inorganic $2 \lambda$ GaAs/AlAs$_{0.98}$P$_{0.02}$ planar microcavity with embedded In$_{0.08}$Ga$_{0.92}$As quantum wells~\cite{cilibrizzi_polariton_2014}. The sample is held in the cryostat at 4K and excited nonresonantly so that two excitation lasers are detuned by $\approx$ 100 meV from the lower polariton branch. The excitation continuous wave (CW) lasers are chopped with an acousto-optic modulator to form 2 $\mu$s pulses to reduce the sample heating. The cavity mode is negatively detuned from the exciton level by $-3$ meV. In order to optically induce the time-periodic rotating potential to polaritons, we utilise the technique described in Ref.~\cite{rotatingbucket}. In brief, two frequency-detuned and wavelength-stabilised single-mode lasers $f_{1,2}$ are shaped with two reflective spatial light modulators (SLMs) in the form of the ``perfect vortex'' beam~\cite{Chen2013_perfectvortex}. Thus, the transverse profile of each beam is a ring with imprinted phase winding. With the SLMs, we independently control both beams' diameter $d$ and optical angular momentum $l$. At fixed frequency difference $\Delta f$ of the lasers, we imprint the $l_{1,2}=\pm1$ for the first and the second laser beam, respectively, or vice versa. Then, the two annular beams are spatially overlapped, forming a profile of broken axial symmetry, and focused onto the sample [see Fig.~\ref{fig1}(a)]. As a result, the profile rotates at the frequency $f$ governed by the lasers' frequency difference $\Delta f$ and their OAM $\Delta l$ difference~\cite{rotatingbucket,rotating}: 
\begin{equation}\label{eq1}
     f = \frac{\Delta f}{\Delta l}=\frac{f_1-f_2}{l_1-l_2}.
\end{equation}
The sign of $f$ defines the rotation direction of the total beam profile with anti-clockwise and clockwise corresponding to positive and negative $f$, respectively.

Alongside stirring, the non-resonant laser pattern also induces a time-average optical trap for the polaritons~\cite{askitopoulos_polariton_2013}. Optically injected background reservoir excitons, co-localized with the beam profile, blueshift polaritons because of the repulsive exciton-exciton (and polariton) Coloumb interaction. As such, the rotating beam profile defines a time-periodic potential landscape for the polaritons [see Fig.~\ref{fig1}(a)], whose size and trap depth define the number of allowed confined energy levels. It is well known that in such systems, the polariton condensation might occur at one of the trap modes at the balance of the polariton gain and dissipation~\cite{askitopoulos_robust_2015}.

For the smallest optical trap with a diameter $d=14$ $\mu$m rotated at $f = $ 2 GHz, we observe a single quantised vortex~\cite{rotatingbucket}, corresponding to the condensate getting stirred into the trap's first excited state manifold [see Fig.~\ref{fig1}(b)]. The condensate real space (near field) phase distribution is measured through a homodyne interferometry technique~\cite{sergey_prl}. When we increase the diameter of the annular excitation profile to $d=15.4$ $\mu$m and $d = 17.5$ $\mu$m, at approximately the same stirring frequency, we observe two and three vortices, respectively, of the same topological charge $l=1$, as depicted in Fig.~\ref{fig1}(c) and~\ref{fig1}(d). The increasing number of vortices with trap size implies that the condensate is shifting its population to higher order trap states as they become available~\cite{Dreisman_PNAS2014, Sun_PRB2018, Ma_PRL2018, Nalitov_PRA2019, Alperin_Optica2021}. Our experimental results are reproduced through numerical simulations, shown in Figs.~\ref{fig1}(e-g), using the generalized Gross-Pitaevskii equation (see Supplemental Material for details).

\begin{figure}[t]
    \centering
    \includegraphics[width=0.99\columnwidth]{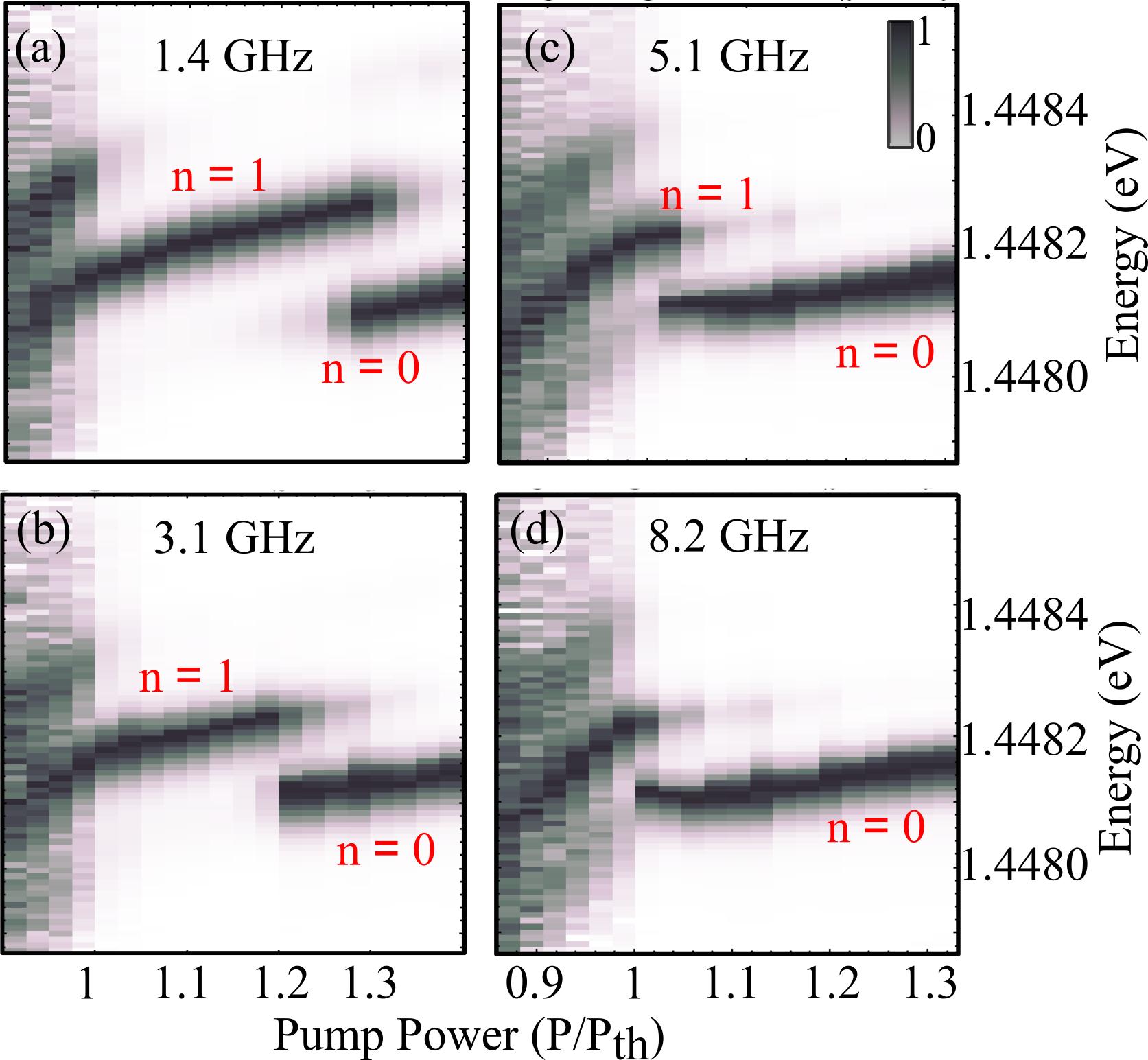}
    \caption{The condensate normalised spectra as a function of pump power for different rotation frequencies (a) $f = $ 1.4 GHz, (b) 3.1 GHz, (c) 5.1 GHz, (d) 8.2 GHz. The excitation trap diameter is fixed at $d= 14$ $\mu$m and $n$ denotes the trap's principle quantum number.}
    \label{fig2}
\end{figure}

To gain a better insight into the physics of the vortex cluster formation, we perform a series of excitation power and rotation frequency scans for each trap size. We start with the smallest trap ($d$ = 14 $\mu$m) containing just a single vortex. 
Figure~\ref{fig2} shows the energy-resolved polariton photoluminescence measured for different rotation frequencies and pump powers. Below the condensation threshold $(P_\text{th})$, the polaritons are broadly distributed across several energy branches corresponding to the confined states of the optical trap. As soon as the threshold is reached, bosonic stimulation forces polaritons to populate a specific energy state accompanied by linewidth narrowing~\cite{kasprzak_bose-einstein_2006}. At low powers, the condensate is first driven into the branch belonging to the first excited state manifold ($n=1$), corresponding to the trap's $l=\pm 1$ angular harmonics. In this branch, the condensate can get stirred into a definite direction within a certain rotation frequency interval $|f| \in [1,4]$ GHz as reported in Ref.~\cite{rotatingbucket}. 

When scanning the pump power, we reveal a transition from the excited to the ground state branch ($n=0$) implying that the condensate vortex eventually destabilises at higher reservoir densities with condensation taking place in the Gaussian ground state. Interestingly, as we increase the rotation frequency from  $f = 1.4$ GHz [Fig.~\ref{fig2}(a)] to $f = 8.2$ GHz [Fig.~\ref{fig2}(d)], we find that the ground state becomes dominant over the range of investigated powers. As the pump rotation speed increases the anisotropic reservoir (trap) profile is smeared out, approaching cylindrical symmetry and restoring the degeneracy between clockwise and anticlockwise polariton currents with no deterministic vorticity forming~\cite{rotatingbucket}. Moreover, when the pump power is increased, interactions between the reservoir and the condensate are enhanced, which facilitates energy relaxation of the condensate~\cite{Wouters_PRB2010, Sun_PRB2018}. When both effects, fast rotation and high power, are sufficiently strong, they drive the condensate from a vortex state into the vortex-free ground state. This is in sharp contrast to conventional rotating superfluids~\cite{hevortex} and BECs~\cite{bec_review} as we do not observe an increasing number of vortices for higher rotation frequencies. However, further we show that for the bigger optical traps, the condensate can sustain a bigger number of vortices.

\begin{figure}[t]
    \centering
    \includegraphics[width=0.95\columnwidth]{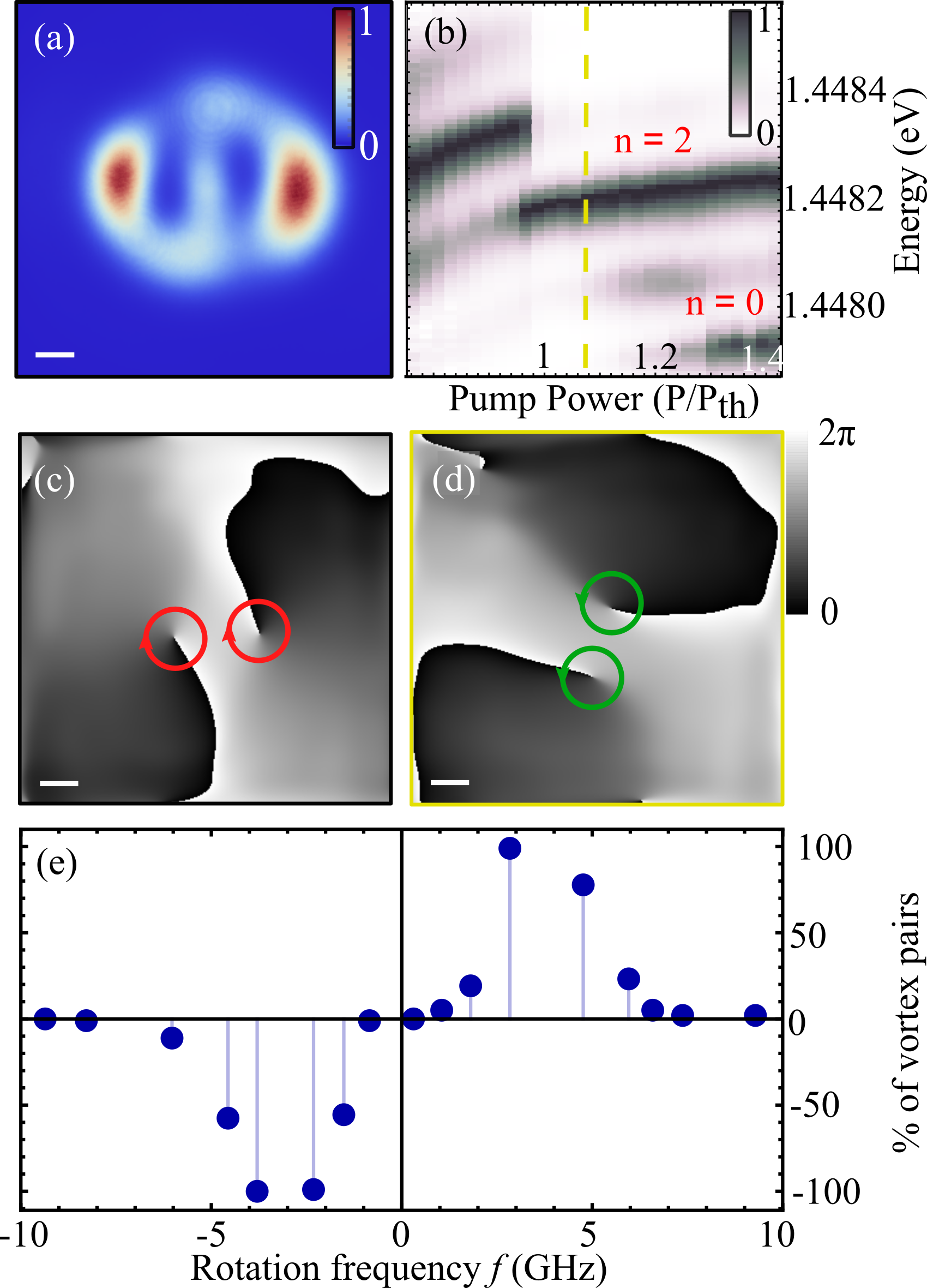}
    \caption{ (a) The real-space condensate intensity distribution for the condensate formed in the 15.4 $\mu$m diameter trap rotating at $f = -3$ GHz. (b) The spectrum power dependence for the 15.4 $\mu$m trap ($f = -3$ GHz). The yellow dashed line depicts the pump power corresponding to the retrieved phase profile in panel (d). Condensate phase distribution for (c) $f = -3$ GHz and (b) $f = 3$ GHz. (e) Experimentally obtained frequency dependence for the vortex pair occurrence. The scale bar in panels (a), (c), (d) corresponds to 2 $\mu$m.  }
    \label{fig3}
\end{figure}

We next study a larger rotating trap with $d = $ 15.4 $\mu$m in which, for $f = 3$ GHz, the condensate intensity distribution [see Fig.~\ref{fig3}(a)] now features two intensity minima and dominantly occupies the second excited state manifold ($n=2$) of the trap above threshold [see Fig.~\ref{fig3}(b)]. Note that the non-uniform intensity distribution around the circumference of the condensate and its slightly elliptical shape is connected to the imperfections of the annular lasers intensity distribution, making the optical trap slightly elliptical. The two intensity minima correspond to two co-rotating vortices with $l =  -1$ revealed in the real-space phase distribution [see Fig.~\ref{fig3}(c)] or $l = +1$ [see Fig.~\ref{fig3}(d)] depending on the trap rotation direction. Note that the alignment of the vortices is determined by the experimental anisotropy in the laser excitation profile. 

%
%
Interestingly, for a similar range of investigated powers as in Fig.~\ref{fig2}, we do not see an abrupt condensate transition into the lower energy branches of the bigger trap in Fig.~\ref{fig3}(b). Instead, above $1.3 P_\text{th}$ the condensate dominantly occupies both the second and ground energy modes. We note that by fine-tuning the size of the excitation potential, it is possible to create one that favours a cascade of single-mode condensations into different trap energy levels with increasing pump power. In Supplementary Material, we analyse such optical trap with diameter $d = 14.7$ $\mu$m that favours the consequent condensation in the first excited and ground state with the growing excitation laser power. The number of stirred vortices is changing on a par going from a pair of co-rotating vortices to a single one. 

Similar to the smaller trap~\cite{rotatingbucket}, the pair of co-rotating vortices in  $d = $ 15.4 $\mu$m trap appear only in a narrow range of stirring frequencies. To reveal this dependence, we analyse the phase distribution of the condensate at different rotation speeds. For each rotation we record 99 excitation shots and obtain corresponding interference patterns of the condensate interfering with the retroreflected displaced copy of itself in a Michelson interferometer. We then extract the azimuthal phase profile at a small fixed distance around each of the two phase dislocations in the condensate to check for vorticity and build the histogram presented in Fig.~\ref{fig3}(e). This graph reflects the percentage of vortex pair occurrences multiplied by the topological charge of the co-rotating vortices ($l =1$ or $-1$) [i.e., normalized average angular momentum in the condensate].  The average angular momentum in Fig.~\ref{fig3}(e) follows the sign of the rotation frequencies $f$ meaning that the pair of vortices co-rotate with the stirring direction. Moreover, the vortices are formed above the critical rotation frequency of $|f|\approx1.5$ GHz and disappear above $|f|\approx 6$ GHz. 
Notice that both critical frequency and the range of appearance are bigger than that of the smaller 14 $\mu$m trap. This is attributed to the increased circumference of the trap, which, on the one hand, requires a bigger rotation frequency to form a confining potential to sustain the second excited state. On the other hand, the bigger size of the excitonic reservoir does not smear out for the faster rotation and still induces the torque to the condensate. It is worth mentioning that, unlike the conventional superfluid~\cite{hevortex,bec_review}, above the critical rotation frequency, the condensate forms in the state with two vortices bypassing the state with one vortex. 

\begin{figure}[t]
    \centering
    \includegraphics[width=0.9\columnwidth]{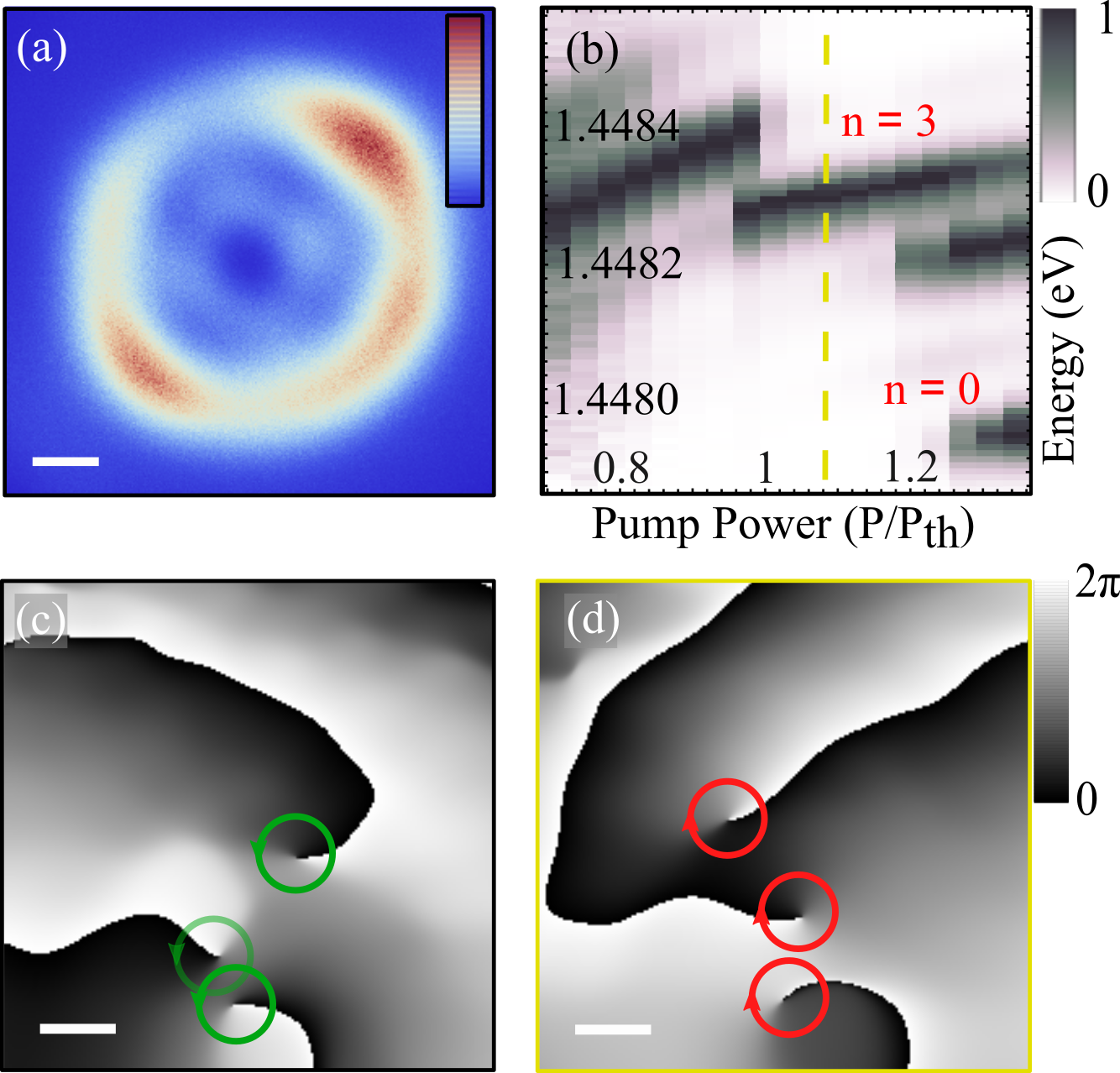}
    \caption{(a) Condensate intensity distribution for the excitation with the rotating trap of $d = 17.5$ $\mu$m. (b) The spectrum power dependence measured for $d = 17.5$ $\mu$m trap rotating at $f=-2$ GHz. The real-space phase maps extracted for the condensate (interfered with the displaced and retro-reflected copy of itself) stirred at (c) $f=$ 2 GHz and in the opposite direction with (d) $f=-2$ GHz.}
    \label{fig4}
\end{figure}

We next increase the trap size even more, to $d = $ 17.5 $\mu$m. The real-space photoluminescence intensity of the condensate in such rotating potential ($f = -2$ GHz) is presented in Fig.~\ref{fig4}(a). The size of the condensate is increased on a par with the optical trap, and just above the threshold, it occupies the third ($n = 3$) exited state manifold of the trap [see Fig.~\ref{fig4}(b)]. Similarly to the 15.4 $\mu$m trap, the condensation is multi-modal for the higher excitation power, with the condensate occupying three energy states above $1.2P_\text{th}$. Retrieving the phase of the condensate at $1.1P_\text{th}$ we now find a cluster of three co-rotating vortices with $l = -1$ [see Fig.~\ref{fig4}(d)]. Inverting the rotation direction to $f= 2$ GHz makes all vortices depicted in Fig.~\ref{fig4}(c)  flip their topological charges to follow the pump rotation. As mentioned before, the arrangement of the vortices in the cluster is dictated by non-homogeneity in the optical potential and cavity disorder. This is also confirmed with the numerical simulation in Figs.~\ref{fig1}(f),(g).

To sum up, we investigated the formation of vortex clusters in nonresonantly optically stirred polariton condensates. We demonstrate that the number of vortices in the clusters depends on the size of the pump-induced rotating trap and the excitation power, with the maximum number achieved in this study being 3 vortices. We also show that the vortices in the cluster deterministically co-rotate with the trap stirring direction despite the lack of any phase-imprinting coming from the excitation beam in sharp contrast to Ref.~\cite{Choi_PRB2022} (i.e., our technique is fully nonresonant). Moreover, we find that the energy spectrum of the condensate is rotation frequency-dependent, which is attributed to the finite lifetime of the photoexcited background exciton reservoir, which forms the optical trap.

Our findings are relevant to the emerging research direction focused on polariton condensates in time-periodic potentials~\cite{rotatingbucket,fraser2023, gnusov_2023optically, yulin2022spin}. The redistribution of energy states in the condensate with the stirring speed opens new prospects for the utilization of polaritons as a platform for Floquet engineering~\cite{floquet_review}. Additionally, advances in all-optical polariton vortex lattices~\cite{sergey_vortex_lattice} in conjunction with our method can be used to locally manipulate vorticity at individual lattice sites. Finally, the achieved control over polariton flow and state can also benefit the realisation of the polariton quantum bits~\cite{Kavokin2022}, as well as sources of structured, coherent light for different applications.


The data presented in this paper are openly
available from the University of Southampton repository [the link is to be provided].

The authors acknowledge the support of the European Union’s Horizon 2020 program, through a FET Open research and innovation action under the grant agreement No. 899141 (PoLLoC) and No.964770 (TopoLight).

H.S. acknowledges the Icelandic Research Fund (Rann\'{i}s) grant No. 239552-051.

\setcounter{equation}{0}
\setcounter{figure}{0}
\setcounter{section}{0}
\renewcommand{\theequation}{S\arabic{equation}}
\renewcommand{\thefigure}{S\arabic{figure}}
\renewcommand{\thesection}{S\arabic{section}}
\onecolumngrid
\newpage
\vspace{1cm}
\begin{center}
\Large \textbf{Supplemental Material}
\end{center}

\section{Numerical Simulations}

A rotating optical trap feeds and confines an excited state condensate, which also rotates with the pump. The effects of this rotating pump profile are modelled using a generalised form of the two-dimensional Gross-Pitaevskii equation coupled to two exciton reservoirs. The condensate wavefunction is described by order parameter $\Psi(\textbf{r},t)$, giving the polariton density. The phase of the condensate can be extracted as the argument of this complex wavefunction, $\arg \Psi(\textbf{r},t)$. The exciton reservoir is made up of two components: the active reservoir, $n_A(\textbf{r},t)$, which contains optically-active excitons that feed the condensate through bosonic stimulation, and the inactive reservoir, $n_I(\textbf{r},t)$, containing the excitons that sustain the active reservoir and is directly excited by the laser pump profile, $P(\textbf{r},t)$. These dynamics are described following:
\begin{gather}
i\hbar\frac{\partial \Psi}{\partial t} = \bigg[-\frac{\hbar^2\nabla^2}{2m}  + G(n_A + n_I) + \alpha\lvert\Psi\rvert^2 + \frac{i\hbar}{2}(Rn_A - \gamma)\bigg]\Psi  \nonumber \\
\frac{\partial n_A}{\partial t} = -(\Gamma_A + R\lvert\Psi\rvert^2)n_A + Wn_I 
\label{eq:rot_GPE}\\
\frac{\partial n_I}{\partial t} = -(\Gamma_I + W)n_I + P(\textbf{r},t). \nonumber
\end{gather}
where $m$ is the effective polariton mass, $G = 2g\lvert X \rvert ^2$ and $\alpha = g\lvert X \rvert ^4$ are the polariton-reservoir and polariton-polariton interactions strengths respectively, where $g$ is the exciton-exciton dipole interaction strength and $\lvert X\rvert^2$ is the excitonic Hopfield coefficient. Additionally, $R$ is the rate of stimulated scattering of polaritons into the condensate from the active reservoir, $\gamma=1/\tau_p$ is the polariton decay rate (inverse of polariton lifetime), $\Gamma_{A,I}$ are the active and inactive reservoir exciton decay rates, $W$ is the inactive to active reservoir exciton conversion rate, and $P(\textbf{r},t)$ describes the nonresonant pumping profile that rotates in time with frequency $f$.

The pump profile is formed through the addition of two Laguerre Gaussian profiles with angular momenta different by $\pm2$ (as in simulation):
\begin{align}
P_{LG}(\mathbf{r},t) & = \mathcal{P}(r) \left\lvert e^{i (l_1 \theta -  \omega_1 t )} + e^{i (l_2 \theta -  \omega_2 t)} \right\rvert^2  \nonumber \\
& = 4 \mathcal{P}(r) \cos^2{\left[ \frac{(l_1-l_2) \theta}{2} - \frac{(\omega_1 - \omega_2) t}{2} \right]}, 
\label{eq.pump}
\end{align}
where $\mathcal{P}(r)$ represents the annular intensity profiles of each Laguerre Gaussian beam.

To reproduce the co-rotating vortices with the rotating trap that are observed in the experiment, an additional elliptical profile, $P_\text{ell} = e^{[(x/a)^2 + (y/b)^2]/2\sigma^2}$ also rotating with frequency $f$, is added to $P_{LG}$. Here, $a>b$ with $a$ chosen to match the diameter of $P_{LG}$. The resulting pump profile is given by:
\begin{align}
    P(\textbf{r},t) = \frac{1}{4}P_{LG}(\textbf{r},t) + P_\text{ell}(\textbf{r},t) 
\end{align}
Equations~\eqref{eq:rot_GPE} are numerically integrated in time using a linear multistep method starting from weak random initial conditions (maximum amplitude $10^{-4}$). The parameters used in these simulations are based on the sample properties~[38], with $m = 5.3 \times 10^{-5}m_0$ where $m_0$ is the free electron mass, $\gamma = \frac{1}{5.5}$~ps$^{-1}$, $g = 1~\upmu$eV~$\upmu$m$^2$, and $\lvert X\rvert^2 = 0.35$. We take $\Gamma_A = \gamma$ due to the fast thermalisation to the exciton background and the nonradiative recombination rate of inactive reservoir excitons to be much smaller than the condensate decay rate with $\Gamma_I =0.01\gamma$. The remaining parameters are enumerated through fitting to experimental data, giving $R = 0.01$~ps$^{-1}$, and $W = 0.05$~ps$^{-1}$. The nonresonant pump drive term $P(\textbf{r},t)$ uses a similar profile as in the experiment.

\section{Transition from a pair to a single vortex in stirred polariton condensate}

\begin{figure}[t]
    \centering
    \includegraphics[width=0.55\columnwidth]{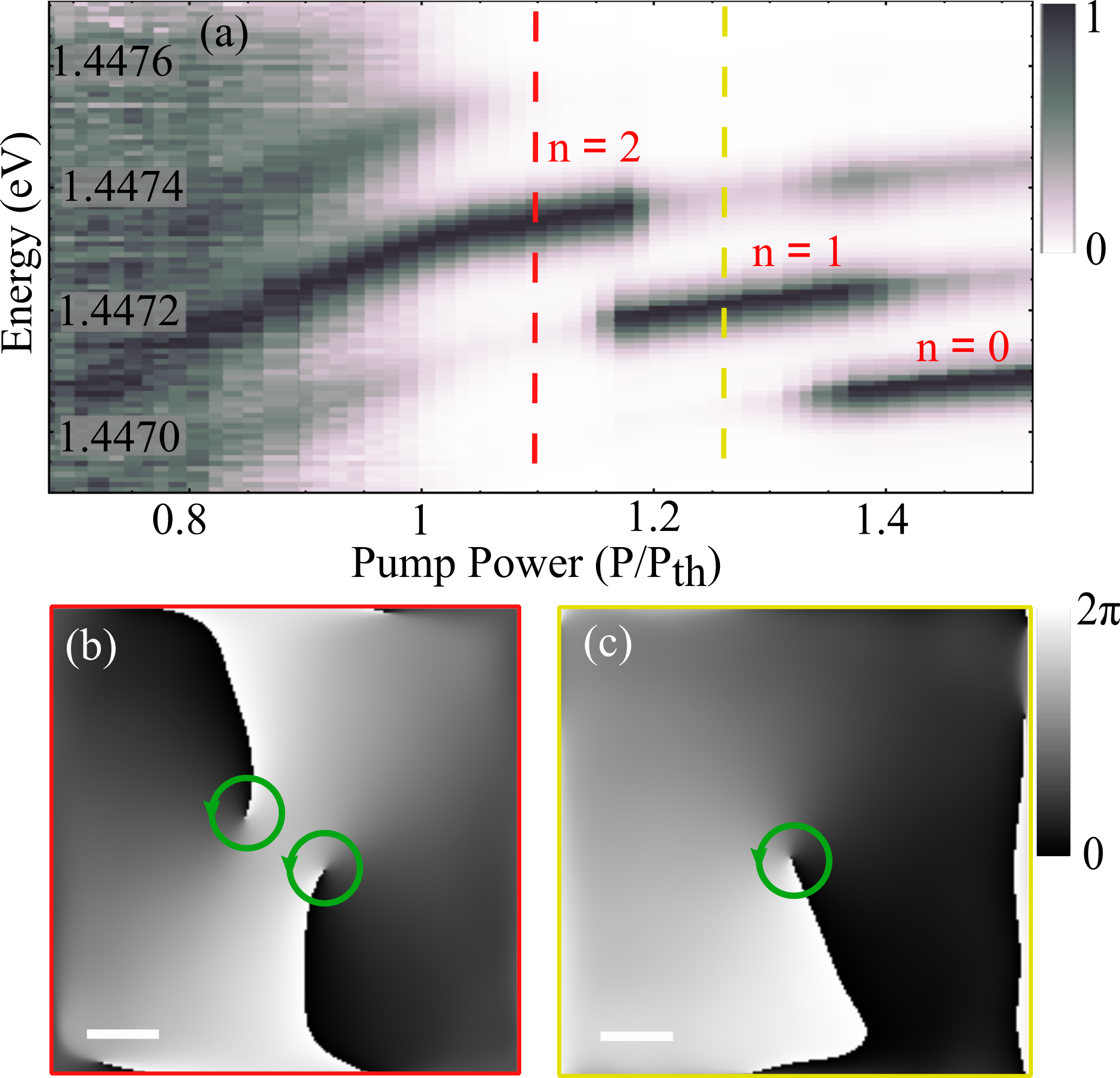}
    \caption{(a) Spectrum of the polariton condensate rotated in an optical trap with a size of $d = $ 14.7 $\mu$m at frequency $f=2.5$ GHz as a function of pump power. The corresponding real-space phase maps of the condensate extracted at pump power (b) $P = 1.1 P_\text{th}$ and (c) $P = 1.25 P_\text{th}$ demonstrating the transition from two vortices to a single one. The red and yellow dashed lines in panel (a) denote the pump power range, corresponding to the extracted phases in (b) and (c), respectively. The scale bar in (b), (c) is 2 $\mu$m.}
    \label{fig5}
\end{figure}

The rotating potentials studied in the main text favoured multi-modal condensation at higher pump powers with a finite population of the condensate occupying the ground state [see Figs.3(b) and~4(b) ]. However, with fine-tuning of  the size of the excitation potential, it is possible to create one that favours a cascade of single-mode condensations into different trap energy levels (see Fig.~\ref{fig5}) with the increasing pump power. Here, we set the trap diameter to $d = 14.7$ $\mu$m (in between Figs.~2 and~3) and set the rotation to $f = 2.5$ GHz. Just above the threshold the condensate occupies the second excited state [see Fig.~\ref{fig5}(a) red dashed line] and holds there until $P = 1.15 P_\text{th}$. With increasing power, the condensate successively starts populating the first exited state (up to $P \approx 1.3P_\text{th}$) and then the ground state. Retrieving the phase distribution of the condensate at different pump power with the homodyne interferometry [see Figs.~\ref{fig5}(b) and~\ref{fig5}(c)], we find that, indeed the condensate vorticity changes. The two co-rotating vortices ($l = 1$) are formed in the second excited state of the confining potential at $P = 1.1P_\text{th}$ and only one vortex of the same topological charge is formed at $P = 1.25P_\text{th}$. Thus, we confirm that the occupied trap energy state defines the vortex number in the rotating confined polariton condensate.  Moreover, this finding facilitates the control over the vortex cluster composition, it shows that the number of vortices in the co-rotating cluster can be controlled not only with the size of the rotating potential but with the excitation power.

\end{document}